\documentclass[journal,twoside,final]{IEEEtran}
\usepackage{amsfonts,amsmath,amssymb,algorithm,algorithmic,bm}
\usepackage{graphicx,epsfig,cite}
\usepackage{color}
\usepackage{subfigure}
\usepackage[percent]{overpic}
\usepackage{multirow}
\usepackage{booktabs}
\usepackage{autobreak}
\usepackage{hyperref}
\usepackage{amsthm}
\usepackage{bbm}
\usepackage{array}
\usepackage{makecell}
\hypersetup{hidelinks}

\def \beqi{\begin{IEEEeqnarray}{rcl}\IEEEyesnumber}
\def \eeqi{\end{IEEEeqnarray}}

\def \bmat{\begin{bmatrix}}
\def \emat{\end{bmatrix}}

\allowdisplaybreaks[3]
\begin{document}

\title{Satellite-Assisted Low-Altitude Economy Networking: Concepts, Applications, and Opportunities}
\author{Shizhao He, Jiacheng Wang, Ying-Chang Liang, {\it Fellow, IEEE}, Geng Sun, {\it Senior Member, IEEE}, and Dusit Niyato, {\it Fellow, IEEE}

\thanks{S. He is with the National Key Laboratory of Wireless Communications, and the Center for Intelligent Networking and Communications (CINC), University of Electronic Science and Technology of China (UESTC), Chengdu 611731, China (e-mail: {heshizhao@std.uestc.edu.cn}).
}
\thanks{J. Wang and D. Niyato are with the College of Computing and Data Science (CCDS), Nanyang Technological University (NTU), Singapore 639798 (e-mail:{jiacheng.wang@ntu.edu.sg;dniyato@ntu.edu.sg})}
\thanks{Y.-C. Liang is with the Center for Intelligent Networking and Communications (CINC), University of Electronic Science and Technology of China (UESTC), Chengdu 611731, China (e-mail: {liangyc@ieee.org}).
}
\thanks{G. Sun is with the College of Computer Science and Technology, Jilin University, Changchun 130012, China (e-mail:sungeng@jlu.edu.cn).}
}

\maketitle
\IEEEpubidadjcol

\begin{abstract}

\textcolor{black}{The low-altitude economy (LAE) is a new economic paradigm that leverages low-altitude vehicles (LAVs) to perform diverse missions across diverse areas. 
To support the operations of LAE, it is essential to establish LAE networks that enable LAV management and communications.}
Existing studies mainly reuse terrestrial networks to construct LAE networks. 
However, the limited coverage of terrestrial networks poses challenges for serving LAVs in remote areas. Besides, efficient LAV operations also require support such as localization and navigation, which terrestrial networks designed for communications cannot fully provide.
Due to ubiquitous coverage and diverse functions, satellites are a promising technology to support LAVs.
\textcolor{black}{Therefore, this article investigates satellite-assisted LAE networking.}
First, we introduce an overview of LAE and satellites, discussing their features, applications, and architectures.
Next, we investigate opportunities for satellites to assist LAE from aspects of communication, control, and computation.
\textcolor{black}{As all assistance depends on reliable satellite-LAV communications, we propose a satellite-assisted LAE framework to tackle issues caused by the severe path loss and high dynamics in satellite-assisted LAE networks.}
{Specifically, the proposed framework comprises distributed satellite MIMO, distributed LAV MIMO, and a two-timescale optimization scheme.
}
\textcolor{black}{The case study demonstrates that the distributed MIMO architecture efficiently reduces the required transmission power and extends service duration, while the two-timescale optimization scheme balances the performance and control signaling overhead.}

\begin{IEEEkeywords}
  Low-altitude economy, Satellite networks, Distributed MIMO.
\end{IEEEkeywords}

\end{abstract}

\vspace{-1em}

\section{Introduction} \label{sec:intro}

The low-altitude economy (LAE) is an emerging economic paradigm operating below $1,000$ m above the ground. In LAE, massive low-altitude aerial vehicles (LAVs), like unmanned aerial vehicles (UAVs) and electric vertical takeoff and landing vehicles (eVTOL), are utilized to perform various missions, such as communication, transportation, and surveillance across diverse areas \cite{jiang20236g}.
The LAE has attracted worldwide attention.
For example, the two largest aerospace companies, Airbus and Boeing, both plan to introduce their aerial taxis\footnote{https://www.boeingfutureofflight.com/wisk}, while some low-altitude transportation projects have also been launched\footnote{https://www.ehangzhou.gov.cn/2024-07/05/c$\_$290091.htm}.
\textcolor{black}{However, to ensure efficient operations of LAE, it is essential to establish LAE networks that enable LAV management and communications.}

The emerging low-altitude applications in LAE impose critical requirements on LAE networks, such as accurate sensing and $3$D coverage.
Recent research utilized terrestrial networks with new technologies, such as massive multi-input multi-output (MIMO), integrated communication and sensing (ISAC), and reconfigurable intelligent surfaces (RIS), to tackle the above issues \cite{jiang20236g}.
Specifically, ISAC enables base stations (BSs) to simultaneously communicate with LAVs and detect their positions.
Besides, massive MIMO and RIS realize $3$D beamforming, allowing BSs to provide $3$D coverage to LAVs.
However, LAE networks built solely on terrestrial networks are insufficient to support LAE due to the limited coverage of BSs \cite{baltaci2021survey}. 
Given the limited coverage, BSs struggle to serve LAVs operating in remote areas. The limited coverage also leads to frequent handovers for highly mobile LAVs, which shortens service duration. Moreover, as BSs are mainly designed for communication,
they are inadequate at providing other essential services, such as precise navigation data \cite{zeng2019accessing}. 


Satellites offer a promising way to overcome the limitations of terrestrial networks. 
Compared to BSs, satellites provide significantly broader coverage. Particularly, the ultra-dense satellite constellation of low earth orbit (LEO) satellites, such as Starlink, enables ubiquitous and high throughput satellite communications.
Besides, satellites are designed with various functions, including communication, navigation, and remote sensing \cite{zeng2019accessing,baltaci2021survey}.
Thus, introducing satellites into LAE networks offers the following advantages:
\begin{itemize}
  \item \textbf{Ubiquitous Coverage:} 
 Ultra-dense satellite constellations provide global coverage to LAVs, ensuring ubiquitous and continuous communication.
  \item \textbf{Comprehensive Support:} Existing satellites can provide diverse services, such as communication and navigation, offering comprehensive support to LAVs.
\end{itemize}

Motivated by the potential benefits of integrating satellites into LAE, this article investigates satellite-assisted LAE networks. 
Achieving satellite-assisted LAE networks depends on reliable satellite-LAV communications, which is challenging due to the severe path loss and high dynamics inherent in such networks. Although satellite communication transceivers effectively mitigate path loss, their high energy consumption makes them unsuitable for energy-limited LAVs \cite{zeng2019accessing}. Besides, high dynamics brings frequent handovers and large signaling overhead. 
To address the above issues, we propose a distributed MIMO based framework for the satellite-assisted LAE network. In the framework, satellites and LAVs within a fleet form distributed MIMO, enabling joint signal processing and introducing higher channel diversity to mitigate interference and extend service duration.
A two-timescale optimization scheme is further applied to adjust beam directions at the frame level and signal processing at the slot level, reducing signaling overhead.
The main contributions are summarized as follows:
\textcolor{black}{\begin{itemize}
  \item We present a general view of the LAE and satellites, including applications, attributes, and existing systems. 
  Based on this, we present the emerging opportunities of satellites to assist LAE networks from the perspectives of communication, control, and computation.
  \item We propose a satellite-assisted LAE network framework, which leverages distributed MIMO to tackle the high energy consumption and frequent handovers, and utilizes a two-timescale optimization scheme to reduce signaling overhead.
  \item We evaluate the proposed framework through a case study, where different LAV fleets upload information to satellites. Numerical results show that the proposed framework effectively reduces energy consumption, extends service duration, and balances the performance and control signaling overhead.
\end{itemize}}




\section{Overview of LAE and Satellites}\label{sec:sysmodel}

\begin{figure*}[htbp]
  \begin{center}
  \includegraphics[width=\textwidth]{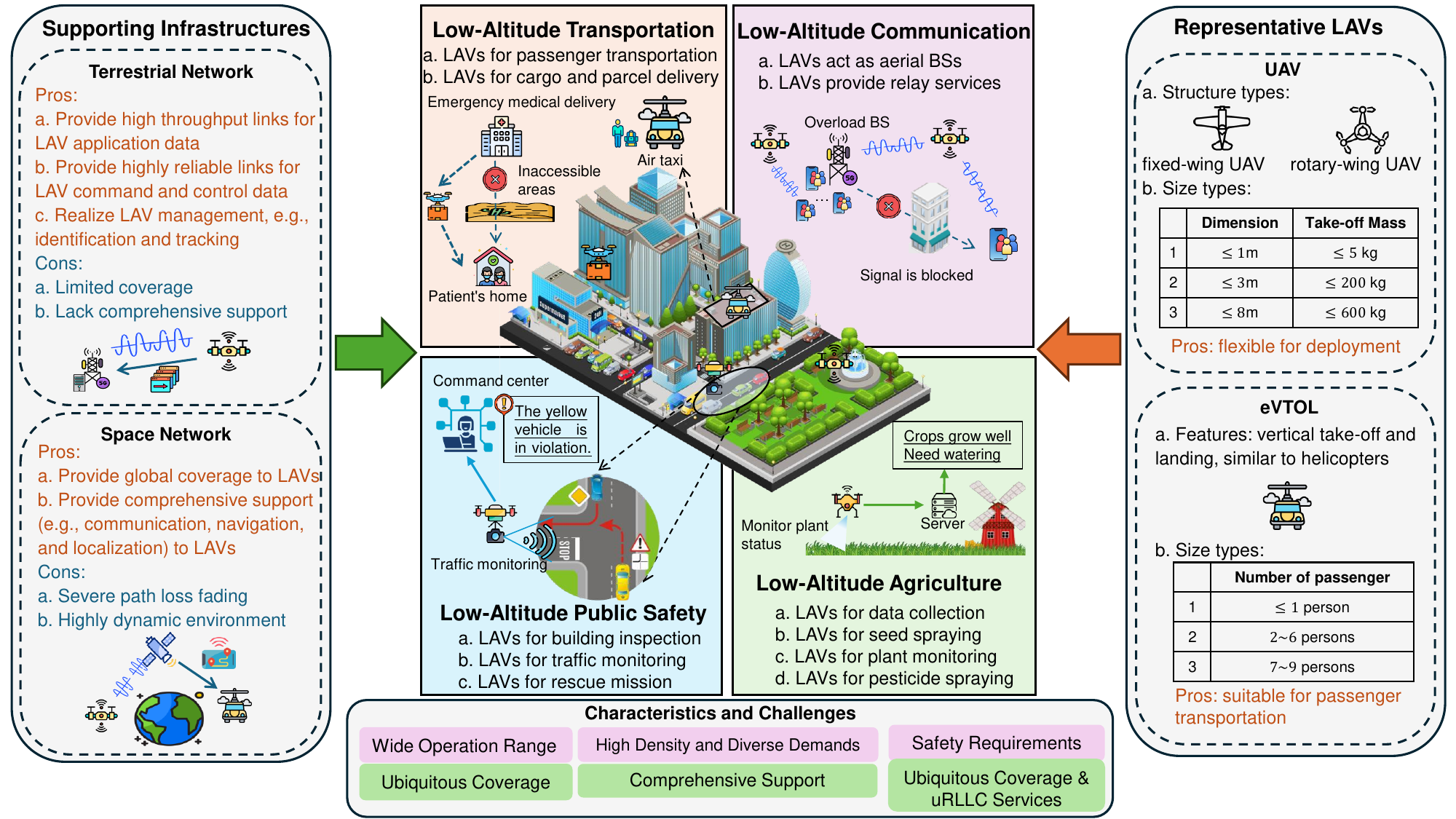}
  \caption{Overview of the LAE networks. We demonstrate four major applications in LAE, including transportation, communication, agriculture, and public safety. 
  To support the mentioned applications, terrestrial and space networks should be utilized to provide comprehensive support to LAVs.
  UAVs and eVTOLs are two representative types of LAVs utilized for different tasks, and we summarize their features and advantages. 
  }\label{fig:sys_model}
  \end{center}
\end{figure*}

This section first introduces the concept of LAE, including representative LAVs, applications, and major features, which are shown in Fig. \ref{fig:sys_model}. Then, we review existing research on LAVs and satellite networks.

\subsection{Low Altitude Economy}

LAE is a new economic paradigm that encompasses a wide range of applications enabled by diverse LAVs. Among them, UAVs and eVTOLs are two representative types:
\begin{itemize}
  \item \textbf{UAV:} UAVs, also known as drones, are categorized into fixed- or rotary-wing types. 
  Within each type, UAVs vary widely in dimension and \textcolor{black}{take-off mass}, based on which they can be classified into \textcolor{black}{three types} \cite{baltaci2021survey}. In general, UAVs are highly flexible for deployment and can be used anywhere for surveillance, communication, and transportation.
  \item \textbf{eVTOL:} eVTOLs are designed for vertical take-off and landing, which are mainly used for passenger transportation within medium ranges ($<80$ km). They can be regarded as a special form of helicopters with electric engines, supporting air shuttles, air taxis, and air ambulances \cite{baltaci2021survey}. 
\end{itemize}

As shown in Fig. \ref{fig:sys_model}, these LAVs can support diverse low-altitude applications, including communication, transportation, agriculture, and public safety.
\begin{itemize}
  \item \textbf{Low-Altitude Communication:} The UAVs can work as aerial BSs or relays to extend the coverage of cellular networks, providing communication services to remote areas and traffic offloading to hotspot areas \cite{zeng2019accessing}.
  Besides, the LAVs can also collect the data of \textcolor{black}{Internet of Things} (IoT) devices located in inaccessible areas.

  \item \textbf{Low-Altitude Transportation:} The eVTOLs can be utilized as air taxis to transport passengers, offering faster and congestion-free transportation. Besides, the large UAVs can be used to transport cargo to remote areas with complex terrain, while the small UAVs can deliver parcels within urban and suburban areas \cite{baltaci2021survey}.
  
  \item \textbf{Low-Altitude Agriculture:} Agriculture is an important application for UAVs, accounting for $80\%$ of the global UAV revenues in 2017 \cite{baltaci2021survey}. UAVs can effectively collect environmental data (e.g., temperature, humidity), monitor the status of plants and animals, and 
  spray seeds and pesticides.
  
  \item \textbf{Low-Altitude Public Safety:} LAVs play a significant role in public safety. For LAVs with surveillance capability, they can inspect infrastructures, such as bridges and pipelines, and monitor transportation traffic. Besides, LAVs can also carry out humanitarian missions, such as searching and rescuing survivors \cite{jiang20236g}.
\end{itemize}


\textcolor{black}{LAE is essentially a new paradigm of UAV networks, extending beyond conventional UAV communication networks where UAVs are limited to communication tasks. In contrast, LAE networks need to accommodate a \textcolor{black}{massive number of} LAVs for different missions, such as transportation and sensing.} Thus, LAE networks present the following characteristics:
\begin{itemize}
  \item \textbf{Wide Operation Range:} 
  LAVs need to support applications across diverse areas, including cities, rural, and post-disaster areas \cite{zeng2019accessing}. 
  This extensive operational range necessitates LAE networks to provide ubiquitous coverage across vast geographical regions.

  \item \textbf{High Density and Diverse Demands:} 
  With UAV deployments expected to exceed $13$ billion by 2030\footnote{https://worldmetrics.org/drones-statistics/}, LAE networks support a high density of LAVs. 
  These LAVs carry out various missions and impose diverse demands on LAE networks, such as precise navigation and low-latency communications \cite{zeng2019accessing,baltaci2021survey}, requiring comprehensive support.
  
  \item \textbf{Stringent Safety Requirements:} 
  The operation of massive LAVs requires stringent safety requirements to prevent collisions and operational hazards.
  The Radio Technical Commission for Aeronautics (RTCA) specifies that the communication availability and continuity for remote piloting are both over $99.9\%$ \cite{baltaci2021survey}.
  This attribute demands ubiquitous coverage and ultra-reliable low-latency communication (uRLLC) connectivity.
\end{itemize}

\subsection{{Existing Research on LAV}}

\textcolor{black}{Existing research mainly focuses on leveraging terrestrial networks, particularly cellular networks, to support LAVs.
The reason is that cellular networks are widely deployed in urban areas and can be reused to support high-capacity and low-latency air-to-ground (ATG) communications \cite{zeng2019accessing}. 
In particular, cellular networks play an important role in LAV management and communications.}
\begin{itemize}
  \item \textbf{LAV Management:} LAV operations are regulated by the unmanned traffic management (UTM) system, which is a set of functions and services, including LAV registration, identification, and tracking \cite{baltaci2021survey}. The Third Generation Partnership Project (3GPP) has developed standards for supporting UTM using cellular networks. For example, TR 23.754 specifies mechanisms for LAV connectivity, identification, and tracking, while TS 22.125 outlines the requirements for remote identification and tracking of LAVs.
  \item \textbf{LAV Communications:} Cellular-connected LAVs have been widely studied in recent years to support LAV communications \cite{zeng2019accessing}. Specifically, 3GPP has released TR 36.777 and TR 22.829 to enhance $4$G and $5$G networks to support LAV communications, respectively.
  Emerging technologies, such as RIS and ISAC, enable cellular networks to achieve $3$D coverage and facilitate both communications and detection \cite{jiang20236g}. \textcolor{black}{Besides, BSs can use deep learning (DL) to predict LAV trajectories and adaptively design beamforming to enhance the reliability of cellular-connected LAV communications \cite{baltaci2021survey}.}
\end{itemize}

Although utilizing cellular networks for LAV management and communications offers promising advantages, it faces several challenges.
{First, cellular networks are mainly deployed in urban areas and lack the infrastructure to provide ubiquitous coverage for LAVs operating in remote areas, such as deserts \cite{baltaci2021survey}. 
Moreover, as cellular networks are mainly designed for high-quality communication services, they have difficulties in adequately addressing other critical LAV requirements, such as high-precision navigation \cite{zeng2019accessing}.}

\subsection{Satellite Networks}

Recent advancements in satellite networks provide a promising solution for establishing LAE. 
Satellite networks comprise spacecrafts orbiting at various altitudes, typically classified into three categories: GEO satellites at $35,786$ km MEO satellites between $2,000 - 20,000$ km, \textcolor{black}{and LEO satellites between $300$-$2,000$ km}. 
Besides, these satellites serve distinct functional purposes, encompassing communication, navigation, and remote sensing  \cite{baltaci2021survey}. 
We summarize several representative satellite networks as follows:
\begin{itemize}
  \item \textbf{Starlink:} Starlink is a communication LEO satellite network, planned to deploy $42,000$ satellites across three orbital altitudes: $340$ km, $550$ km, and $1,110$ km. Starlink aims to provide worldwide Internet services with rates of $17 - 23$ Gbps.
  \item \textbf{Global Positioning System:} The Global Positioning System (GPS) is a well-known satellite navigation system, comprising over $30$ MEO satellites. These satellites broadcast time and location signals, enabling users to realize localization and navigation.
  \item \textbf{Terra and Aqua Satellites:} The Terra and Aqua satellites are launched by NASA for earth observation. These LEO satellites orbit at about $705$ km and employ sensors to monitor environments, such as water cycles and temperature distributions.
\end{itemize}


Compared with terrestrial networks, satellite networks can provide ubiquitous and comprehensive services to users. 
According to 3GPP TR 38.821, three architectures are defined for LAVs to access satellite networks.
\begin{itemize}
  \item \textbf{Transparent Architecture:} Satellites act as relays, simply forwarding signals between ground stations and LAVs without signal processing functions. 
  \item \textbf{Onboard gNB-DU Regenerative Architecture:} Satellites are equipped with a generation nodeB distributed unit (gNB-DU) and able to perform partial BS functions, such as signal processing. 
  These satellites need to connect to a gNB centralized unit (gNB-CU) on the ground for other functions, such as radio resource control.
  \item \textbf{Onboard gNB Regenerative Architecture:} Satellites incorporate complete gNB functionality, operating as BSs by processing signals, managing LAV connections, and performing onboard resource allocation.
  
\end{itemize}

\subsection{\textcolor{black}{Summary}}

Based on the overview of the LAE and satellite networks, we highlight the following critical insights.
\begin{itemize}
  \item \textbf{LAE requires comprehensive support:} The LAE network is a complex system comprising various devices and activities, requiring ubiquitous coverage and comprehensive support across communication and navigation to ensure operational safety and efficiency. 
  \item \textcolor{black}{\textbf{Terrestrial networks in LAE:} Reusing existing terrestrial networks is a cost-efficient method to provide high-quality ATG communications and support LAV management. However, the constrained coverage of terrestrial BSs impedes the ubiquitous coverage. 
	Moreover, BSs designed for communications face challenges in satisfying the comprehensive requirements of LAVs, such as providing precise navigation services to LAVs.} 
  \item \textbf{Integration of satellites and LAE:} With their ubiquitous coverage and multi-functionality, satellites are essential for providing global and comprehensive services to LAVs. 
  Specifically, accessing satellite networks enables LAVs to receive application and command \& communication data anywhere, effectively guaranteeing reliable operations in remote areas \cite{zeng2019accessing}.

\end{itemize}

\section{Satellite Assistance in LAE Networking}\label{sec:com_sagin}

\begin{figure*}[htbp]
  \begin{center}
  \includegraphics[width=\textwidth]{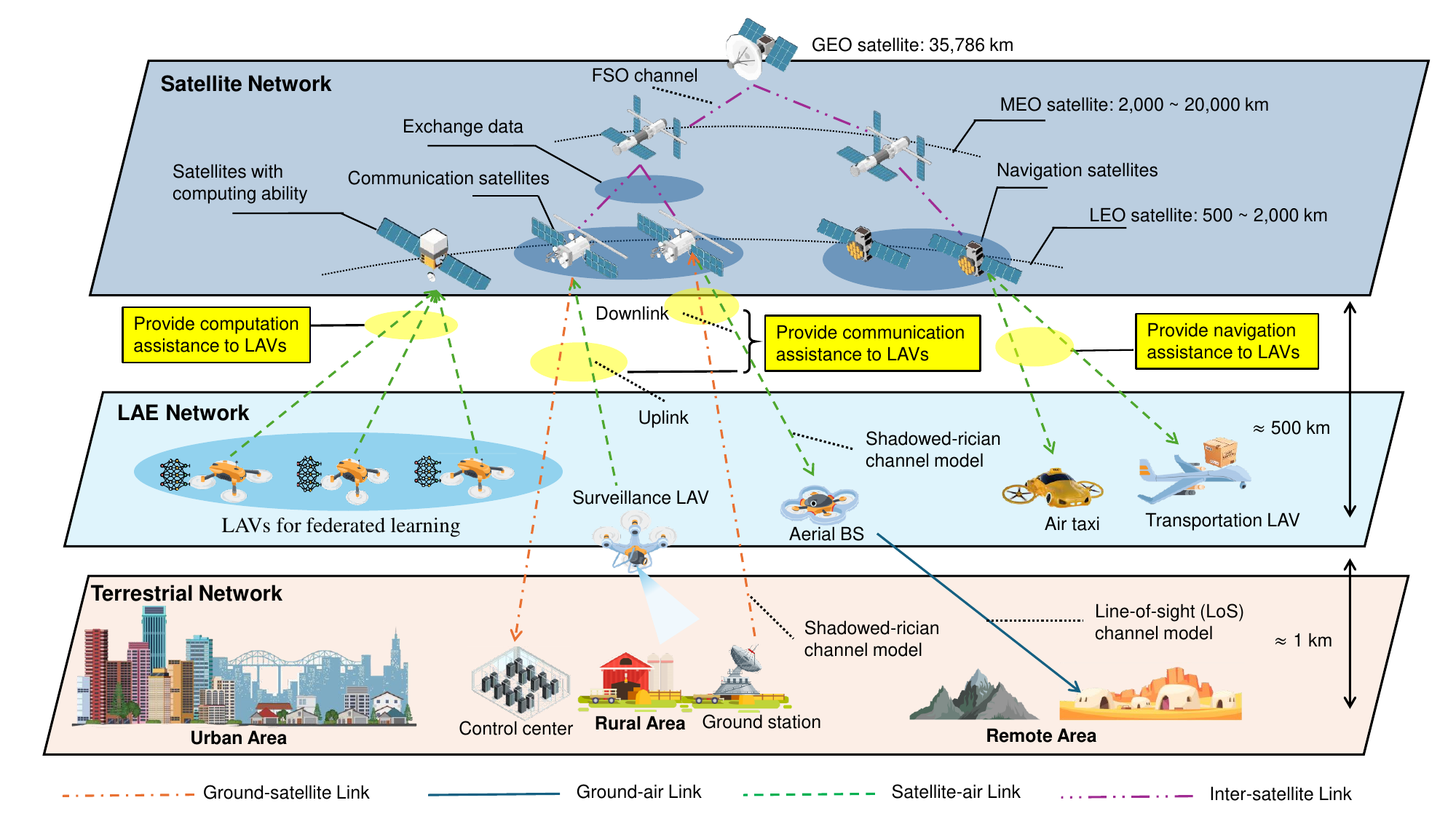}
  \caption{Overview of the satellite-assisted LAE network.
  The satellite network consists of various satellites, including communication satellites, navigation satellites, and computing satellites. Communication satellites provide both downlink and uplink links to LAVs, while navigation satellites offer navigation assistance to LAVs for transportation. Moreover, computing satellites can assist LAVs in FL and task offloading.
  }\label{fig:sat_lae}
  \end{center}
\end{figure*}

In this section, we investigate how satellites can enhance the LAE networks from communication, control, and computation perspectives. The overview of the satellite-assisted LAE network is demonstrated in \textcolor{black}{Fig. \ref{fig:sat_lae}.}

\subsection{Communication Assistance}\label{subsec:rm_se}

Satellites can support LAE networks from the aspect of communication, ranging from multicast services to access services.

\textit{1) \textbf{Multicast Services:}} In LAE networks, LAVs operating in different regions may request the same information, such as the weather condition. However, since LAVs distribute across vast areas, terrestrial networks are difficult to efficiently broadcast or multicast the information to all LAVs due to their limited coverage. 
Satellites are an alternative solution that can efficiently transmit data to large areas due to their wide coverage.
In \cite{zhu2018cooperative}, a satellite collaborates with terrestrial BSs to provide multicast services to users. Specifically, the satellite complements BSs by serving users with poor terrestrial channels. Benefiting from the top-down nature of satellites, the proposed scheme can achieve about a $530\%$ improvement in max-min capacity compared to the BS-only scenario. 


\textit{2) \textbf{Access Services:}} In LAE networks, LAVs may perform tasks in remote areas, such as collecting environmental data. 
However, complex terrain makes it difficult for terrestrial networks to provide access services to LAVs operating in these areas.
Accessing satellite networks is an efficient solution for connecting remote users \cite{he2024toward}.
In \cite{hu2020joint}, the authors considered a system where one LEO satellite assists BSs in providing wireless backhaul links to UAVs serving remote users. The authors optimized user association and resource allocation to meet user demands while minimizing energy consumption. The numerical results show that the proposed scheme can realize almost $1.5$ times the sum rate compared to the scenario without satellites.

\subsection{Control Assistance}

Satellites can provide necessary control information to LAVs, such as position and environmental conditions, enhancing the stability, reliability, and safety of LAV operations.

\textit{1) \textbf{High Precision Navigation:}} Effective navigation is a critical foundation for LAVs to plan safe flight routes. 
Global navigation satellite systems (GNSS) are the primary technology for aerial vehicle navigation due to their worldwide coverage and superior positioning accuracy \cite{wang2025toward}.
In \cite{tang2022gnss}, the authors utilized
BS-based and inertial navigation to supplement satellite navigation, 
enhancing GNSS performance in urban canyon environments. Specifically, a fusion algorithm based on information geometry to fuse multisource navigation data was proposed, combining the advantages of different technologies. Simulation results indicate that the proposed scheme realizes the lowest localization errors, with an average error of about $0.5$ m and a root mean square error of  $0.98$ m.

\textit{2) \textbf{Global Situation Awareness:}} Situation awareness, which acquires information on environment and system conditions, is crucial for efficient and safe LAV operations \cite{baltaci2021survey}. 
Earth observation satellites offer a feasible solution to global situation awareness via capturing images from space, which can be used to realize environmental monitoring \cite{leyva2023satellite}. However, transmitting these images requires high power. To overcome this issue, the authors propose a real-time image compression framework. Numerical results show a $90\%$ reduction in energy usage compared to raw image transmission.


\subsection{Computation Assistance}

Benefiting from the wide coverage, satellites can also help LAVs implement computation tasks.

\textit{1) \textbf{Task Offloading:}} In LAE, LAVs are expected to support diverse applications, such as mobile edge computing \cite{zeng2019accessing}. However, given the limited onboard computing ability, it is difficult for LAVs to process all tasks. An alternative method is to offload them to edge servers or cloud centers with the help of satellites. In \cite{han2023two}, \textcolor{black}{a LEO} satellite constellation is used to assist BSs in offloading tasks from remote users to the core network. Specifically, the authors minimized the total delay by optimizing resource allocation and offloading decisions. Numerical results show that the proposed scheme reduces the average delay by $75.68\%$ compared to the terrestrial network-only scenario.

\textit{2) \textbf{Federated Learning:}} 
Federated learning (FL) enables multiple LAVs to collaboratively train a shared model in a privacy-preserving manner \cite{fang2022olive}. 
As LAVs distribute across different areas, satellites are suitable for acting as the FL server to realize model aggregation and distribution. In \cite{fang2022olive}, the authors use a LEO satellite constellation to aggregate models from geographically distributed devices. Specifically, each satellite first aggregates local models from devices within its coverage and then exchanges these models with others for inter-satellite model synchronization. As model synchronization introduces a high multi-hop communication cost, the authors proposed a framework to enable fast model synchronization.
Simulation results show that the proposed scheme reduces more than $14\%$ training time with only $0.5\%$ accuracy degradation.

\begin{figure*}[htbp]
  \begin{center}
  \includegraphics[width=\textwidth]{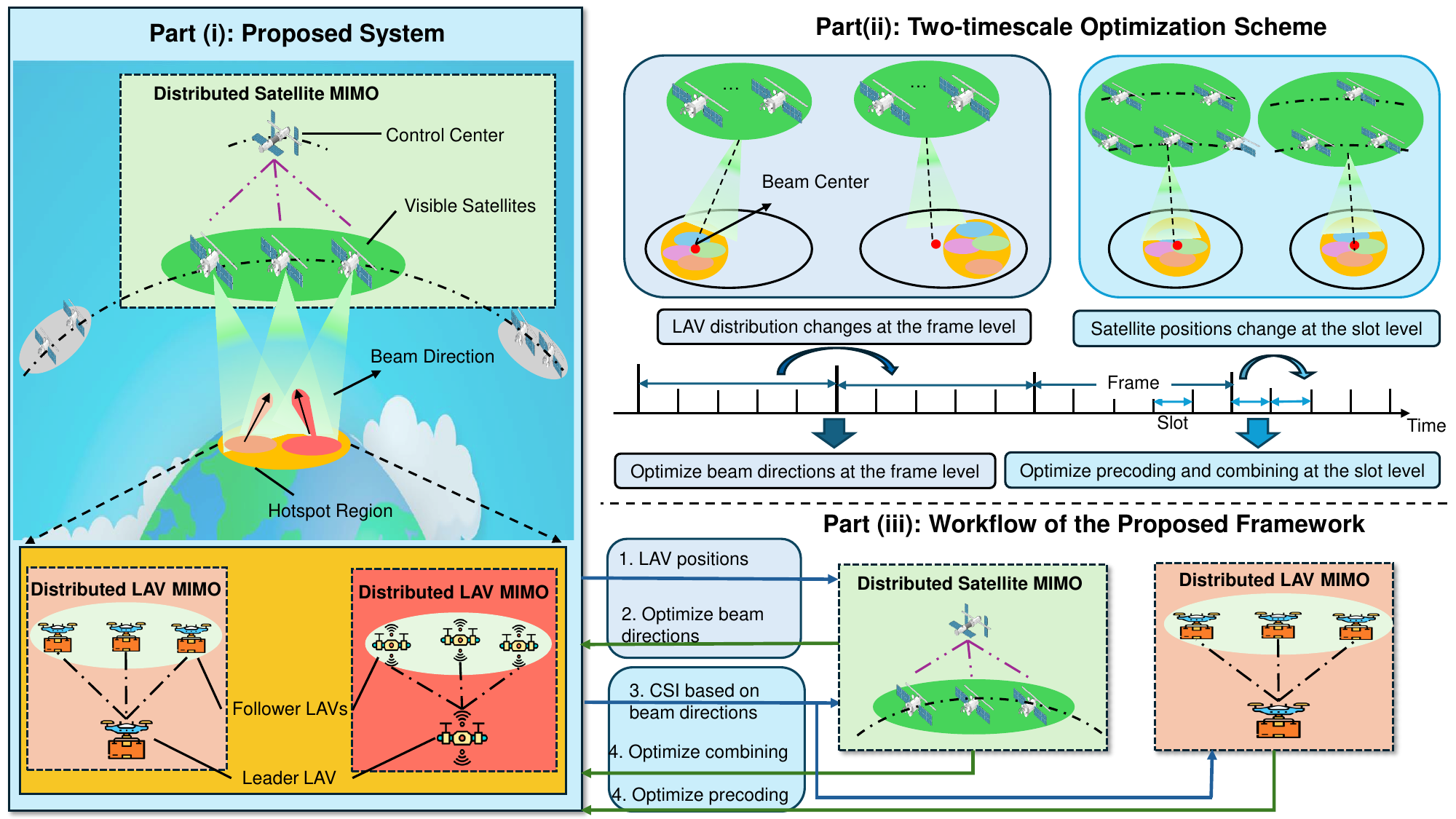}
  \caption{\textcolor{black}{Overview of the proposed framework. 
  \textit{Part (i)} presents the system architecture, where distributed satellite MIMO formed by visible satellites serves multiple distributed LAV MIMO systems, each corresponding to a different LAV fleet.
  \textit{Part (ii)} demonstrates the proposed two-timescale optimization scheme, which optimizes satellite beam directions at the frame level and updates precoding and combining at the time slot level.
  \textit{Part (iii)} shows the operational workflow. At the beginning of each frame, distributed satellite MIMO first acquires LAV positions to determine its beam direction. Then, distributed satellite MIMO and distributed LAV MIMO optimize combining and precoding based on the obtained CSI at the time slot level.}}\label{fig:sat_lae_cf}
  \end{center}
\end{figure*}

\subsection{\textcolor{black}{Summary}}
The discussion of satellite network assistance provides the following key insights.
\begin{itemize}
  \item \textbf{{Comprehensive Satellite Assistance for LAE:}} \textcolor{black}{
  The multi-functionality satellite networks can provide communication, control, and computation assistance to LAVs, enhancing the scalability, reliability, and efficiency of LAE networks.}
  \item \textbf{Communication as the Core of Satellite-Assisted LAE:} {As all assistance depends on reliable data transmission between satellites and LAVs, a well-designed communication framework is fundamental for achieving satellite-assisted LAE.
  }
  \item \textbf{Challenges in Satellite-LAV Communications:} Severe path loss, various interference, and high dynamics are fundamental challenges in satellite-LAV communications. \textcolor{black}{First, the large propagation distances in satellite-LAV communications lead to severe path loss, exceeding $200$ dB at $20$ GHZ\footnote{3GPP TR 38.821}.
  Second, the ultra-dense satellites and their wide coverage cause serious interference, including inter-satellite interference in the downlink and inter-user interference in the uplink \cite{heo2023mimo}.
  }
  Lastly, high orbital velocities of satellites and mobility of LAVs lead to frequent handovers, shortening the service duration.
\end{itemize}

\section{Satellite-Assisted LAE}

This section presents a satellite-assisted LAE framework, where satellites and LAVs leverage the distributed MIMO to overcome challenges in satellite-LAV communications. A two-timescale optimization scheme is integrated to reduce control signaling overhead.
After that, we evaluate the proposed framework through a case study.

\subsection{Motivation}\label{subsec:sat_lae_framework_mot}

Establishing LAV-satellite communication is essential to realizing satellite-assisted LAE networks. 
Most existing works study the downlink communication from satellites to LAVs, however, the uplink communication is more important in LAE applications.
For instance, LAVs for surveillance require uplink data rates of $10$ Mbps, while downlink operations need only $300$ Kbps \cite{zeng2019accessing}. 
Besides high data rates, it is also vital to ensure the service continuity \cite{baltaci2021survey}.
To support high data-rate and reliable uplink communication, the following \textcolor{black}{issues need to be addressed.}
\begin{itemize}
  \item \textbf{High Energy Consumption:} 
  Realizing high uplink data rates requires satellite communication transceivers (e.g., dish antennas) to overcome the severe path loss.
  \textcolor{black}{However, it is impractical for LAVs to carry such power-intensive transceivers due to their limited payload.} Moreover, the high energy consumption of dish antennas would significantly shorten the flight duration of LAVs \cite{zeng2019accessing}.
  \item \textcolor{black}{\textbf{Large Control overhead:} In satellite-LAV communications, satellites need to steer beams toward LAVs to maximize signal strength \cite{yuan2024cache}. This requires high-mobility LAVs to frequently report their positions via the automatic dependent surveillance-broadcast (ADS-B) \cite{baltaci2021survey}, introducing large signaling overhead.}
  
  \item \textbf{Frequent Handover:} LEO satellites orbit at high speeds around $7.5$ km/s, leading to short service duration and frequent handovers \cite{baltaci2021survey}.
  The frequent handovers affect the continuity of services and also increase the signaling overhead.
  
\end{itemize}

\subsection{Proposed Framework}\label{subsec:sat_lae_framework}


To tackle the aforementioned issues, we propose a satellite-assisted LAE framework based on distributed MIMO, which is demonstrated in \textit{Part (i)} of Fig. \ref{fig:sat_lae_cf}. Specifically, there are three important components in the considered framework: distributed satellite MIMO, distributed LAV MIMO, and a two-timescale optimization scheme.

\textcolor{black}{\textit{1) Distributed MIMO:} MIMO is a promising technology to support satellite-LAV communications, as it enhances data rates and reliability under severe path loss conditions \cite{heo2023mimo}. MIMO is generally classified into two paradigms: co-located MIMO and distributed MIMO. The co-located MIMO, where each access point, such as BS, is equipped with multiple antennas to serve its users, faces inter-cell interference. 
Distributed MIMO tackles the above issues by enabling spatially separated APs to cooperatively process signals, introducing higher channel diversity and enhancing interference mitigation \cite{heo2023mimo}.
Besides, given the energy constraints of both satellites and LAVs, deploying large antenna arrays is challenging. 
Thus, the distributed MIMO is more effective for supporting satellite-LAV communications.}

\textit{2) Distributed Satellite MIMO:} 
Ultra-dense LEO constellations enable satellites to coordinate as a distributed MIMO, \textcolor{black}{i.e., distributed satellite MIMO \cite{omid2023space}. 
Existing works} demonstrate that the satellite MIMO can enhance communication performance across downlink, uplink, and handover.
However, these works primarily focus on \textcolor{black}{digital signal processing of satellites (e.g., precoding and combining)}, overlooking the influence of beam directions on antenna gain.
Adjusting beam directions can steer the majority of signals to an intended direction, providing high antenna gain \cite{yuan2024cache}.
To fully harness the potential of satellites, we propose a novel distributed satellite MIMO system that optimizes both digital processing and beam directions.
The proposed satellite hybrid MIMO consists of a control center and a cluster of visible LEO satellites:
\begin{itemize}
  \item \textbf{Control Center:} The control center adopts the onboard gNB architecture, managing beam directions, digital signal processing, and radio resource allocation of LEO satellites.
  \item \textbf{Cluster of Visible LEO Satellites:} 
  LEO satellites visible to LAVs are selected to act as distributed antennas to receive LAV signals.
  Specifically, these satellites adopt the onboard gNB or gNB-DU architecture, following control center instructions to steer beams and design combining.
\end{itemize}

\textit{3) Distributed LAV MIMO:} 
Considering that LAVs within a fleet undertake an identical mission, they can operate as a distributed MIMO system to transmit and receive the same information, referred to as distributed LAV MIMO.
As shown in \textit{Part (i)} of Fig. \ref{fig:sat_lae_cf}, LAVs are classified as \textcolor{black}{the leader LAV and follower LAVs}. 
Similar to the distributed satellite MIMO, the leader LAV functions as the control center for joint signal processing, while follower LAVs operate as distributed antennas to transmit signals \cite{zeng2019accessing}.

\begin{figure*}[htbp]
  \begin{center}
  \includegraphics[width=\textwidth]{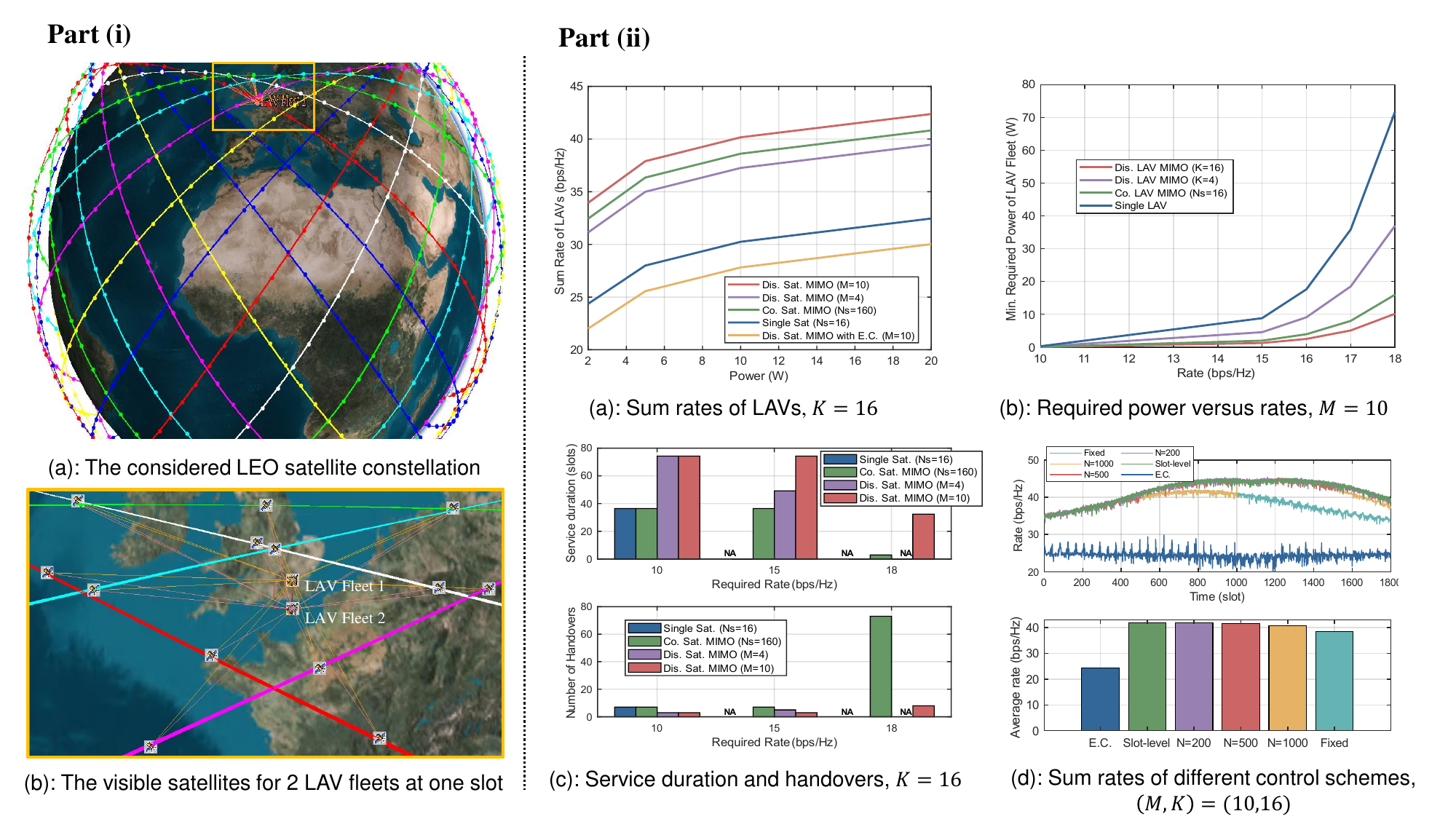}
  \caption{{
  \textit{Part (i)} demonstrates the considered LEO satellite constellation and the visible satellites for both LAV fleets at one slot.
  \textit{Part (ii)-(a)} presents the sum rate of all LAVs versus the transmission power of LAV fleets.
  \textcolor{black}{\textit{Part (ii)-(b)} demonstrates the required transmission power of LAV fleets for achieving target rates.}
  \textcolor{black}{\textit{Part (ii)-(c)} shows the service duration and handovers under different target rates.}
  \textit{Part (ii)-(d)} presents the sum rates during the considered time horizon for different optimization schemes.}
  }\label{fig:result1}
  \end{center}
  \vspace{-2em}
\end{figure*}

\textit{4) Two-timescale Optimization Scheme:} 
As shown in \textit{Part (ii)} of Fig. \ref{fig:sat_lae_cf}, we propose a two-timescale optimization scheme that exploits the difference between LAV movement and LEO satellite dynamics. 
Given high satellite altitudes and the relatively low LAV speeds\footnote{The horizontal speed for aerial vehicles is about $160$ km/h \cite{zeng2019accessing}.}, the fleet distribution appears quasi-static within a frame, resulting in consistent hotspot regions. Conversely, LEO satellites experience substantial position changes over a slot due to their high orbital velocities.
Based on this frame structure, we optimize different resources at two timescales:
\begin{itemize}
  \item \textbf{Frame Level:} Beam directions of distributed satellite MIMO are optimized at the frame level, aligned with the quasi-static hotspot region.
  \item \textbf{Slot Level:} Precoding at LAV MIMO and combining at satellite MIMO are optimized at each slot to adapt to fast-changing channel conditions.
\end{itemize}


\textcolor{black}{
The operation of the proposed framework is illustrated in \textit{Part (iii)} of Fig. \ref{fig:sat_lae_cf} and proceeds as follows. At the beginning of each frame, satellites acquire LAV positions to adjust their beam directions by determining a beam center. 
Then, multiple visible satellites are selected to form the distributed satellite MIMO for each slot. 
Finally, distributed LAV MIMO and distributed satellite MIMO optimize their precoding and combining for uplink transmission.}

The proposed framework can efficiently address the challenges mentioned in Section \ref{subsec:sat_lae_framework_mot}. \textcolor{black}{Specifically, the distributed satellite/LAV MIMO architecture introduces high channel diversity, enabling joint precoding and combining to suppress inter-fleet interference and coherently combine received signals. This enhances the signal-to-interference-plus-noise ratio (SINR) and extends service duration.} 
Moreover, the two-timescale optimization scheme requires LAVs to report positions only once per frame, reducing signaling overhead.

\subsection{Case Study}

\textit{1) Experimental Configurations:} 
\textcolor{black}{In the case study, we leverage the proposed framework to support uplink transmission in a satellite-assisted LAE network, aiming to address the challenges outlined in Section \ref{subsec:sat_lae_framework_mot}.} Specifically, $2$ LAV fleets, each comprising $K$ single-antenna LAVs led by a leader LAV, transmit data to satellites.
As illustrated in \textit{Part (i)-(b)} of Fig. \ref{fig:result1}, LAV fleet $1$ traverses from $(51.48^\circ, -0.076^\circ)$ to $(50.48^\circ, -1.076^\circ)$, while LAV fleet $2$ moves from $(51.48^\circ, -1.076^\circ)$ to $(50.48^\circ, -0.076^\circ)$\footnote{\textcolor{black}{As LAV fleets will traverse ocean during their flights, 
this case reflects practical scenarios such as surveillance tasks in areas lacking BS coverage.}}, both at an altitude of $1$ km and a velocity $v_{\text{LAV}}=0.03$ km/s.
\textcolor{black}{\textit{Part (i)-(a)} of Fig. \ref{fig:result1} demonstrates the considered LEO satellite constellation, which has $22$ circular orbital planes, each with $72$ satellites at an altitude of $550$ km \cite{omid2023space}. The constellation is constructed using the AGI System Tool Kit (STK) with orbital parameters following \cite{omid2023space}. }
Each satellite is equipped with $N_s=16$ antennas, \textcolor{black}{with radiation pattern gains based on ITU-R S.1528 and a half-beamwidth $\varphi_{3\text{dB}}=1.14^\circ$.}
\textcolor{black}{Free-space path loss and Rician fading are adopted to model large-scale and small-scale channel effects, respectively \cite{omid2023space}.}
The considered time horizon is from 10 Dec 2024 04:00 UTCG to 10 Dec 2024 05:00 UTCG, with a slot duration of $2$ seconds and frame length of $N$ slots. The minimum elevation angle required for LAV-satellite communication is $30^\circ$ \cite{omid2023space}, ensuring that at least $10$ satellites are visible to both LAV fleets during the entire period.
\textcolor{black}{Thus, the distributed satellite MIMO can communicate with any LAVs, covering the entire hotspot region. \textit{Part (i)-(b)} of Fig. \ref{fig:result1} demonstrates the visible satellites for both LAV fleets at one slot.}
In this case study, satellites set their beam centers as the centroid of all LAVs. Then, $M$ satellites nearest to the center form the distributed satellite MIMO, processing received signals with the minimum mean square error (MMSE) combiner. Meanwhile, distributed LAV MIMO adopts the maximum ratio transmission (MRT) precoder to transmit its signal.
To demonstrate the performance of the proposed scheme, we consider following baselines. 
\begin{itemize}
  \item \textit{Single Node}: A single satellite/LAV performs reception/transmission, labeled as `single Sat.'/`single LAV'.
  \item \textcolor{black}{\textit{Co-located MIMO}: The distributed satellite MIMO comprising $M$ satellites is replaced by the $MN_s$-antenna satellite with the best channel conditions (`Co. Sat. MIMO'), while the distributed LAV MIMO consisting of $K$ LAVs is substituted by the leader LAV with $Nt=K$ antennas (`Co. LAV MIMO').}
  \item \textit{Earth center beam direction}: Satellites do not optimize their beam directions and always point toward the Earth center, labeled as `E.C.'.
  \item \textit{Fixed beam direction}: The distributed satellite MIMO steers beams toward the initial hotspot region and does not update beam directions. 
  \item \textit{Slot-level beam direction}: The distributed satellite MIMO updates directions at every slot.
\end{itemize}

\textit{2) Performance Analysis:} 
\textit{Part (ii)-(a)} of Fig. \ref{fig:result1} demonstrates the \textcolor{black}{sum of uplink rates of all LAVs}.
{The distributed satellite MIMO improves the sum rate by over $30\%$ compared to the single satellite and outperforms the co-located satellite MIMO at the power of $20$ W. This gain is attributed to higher channel diversity introduced by distributed MIMO, which improves the ability to mitigate inter-fleet interference.}
Distributed satellite MIMO also outperforms the Earth center scheme (`Dis. Sat. MIMO with E.C.'), highlighting the importance of optimizing beam directions. 
On the other hand, \textit{Part (ii)-(b)} illustrates the minimum required power of LAVs for achieving different rates. Distributed LAV MIMO significantly reduces power requirements. {Specifically, to achieve $18$ bps/Hz, the required power is reduced by over $85\%$ compared to the single LAV and is lower than that of the co-located LAV MIMO. }
\textcolor{black}{The above results demonstrate that distributed satellite MIMO and distributed LAV MIMO enable higher rates with lower power, enhancing energy efficiency.}



\textcolor{black}{Next, \textit{Part (ii)-(c)} of Fig. \ref{fig:result1} presents the service duration and handovers of different schemes within a frame comprising $300$ slots. Specifically, LAVs switch to other distributed satellite MIMO or individual satellites when the required rate cannot be maintained. If a scheme fails to meet the rate requirement, it is marked as `NA' in \textit{Part (ii)-(c)}. The top subfigure shows the service duration, while the bottom subfigure illustrates the number of handovers. The distributed satellite MIMO consistently achieves the longest service duration and fewest handovers. Notably, it extends the service duration to $9$ times that of the co-located satellite MIMO at a rate of $18$ bps/Hz. These results confirm that distributed satellite MIMO leverages higher channel diversity to enhance service continuity.}

Lastly, we compare the performance of the proposed two-timescale optimization scheme with other beam direction control strategies in \textit{Part (ii)-(d)} of Fig. \ref{fig:result1}. 
Specifically, we consider three values of $N$ in the proposed scheme, labeled as `$N=200$', `$N=500$', and `$N=1000$', respectively.
The top subfigure shows the sum rate of LAVs over time, while the bottom subfigure presents the average rate of different schemes.
Results show that the proposed schemes with $N=200$ and $N=500$ realize performance comparable to the optimal slot-level scheme and outperform other schemes. Note that a larger $N$ corresponds to fewer signaling overhead for controlling beam directions, and thus the proposed scheme effectively balances performance and signaling overhead. 
These results validate the effectiveness of the proposed two-timescale optimization and prove that treating LAVs as static within $N$ slots is reasonable.

\section{Future Research Directions}\label{sec:future_directions}

\subsection{Intelligence-Empowered Satellite-Assisted LAE}
The high dynamics of satellite-assisted LAE networks pose significant challenges to conventional methods, particularly in handling outdated CSI and enabling real-time decision-making.
Future works should explore artificial intelligence (AI) technologies, including DL and deep reinforcement learning (DRL), to tackle the above issues. In particular, \textcolor{black}{DL can be used to predict CSI based on the location-specific channel knowledge map (CKM), which captures spatial channel features from known satellite and LAV trajectories}. DRL, on the other hand, enables adaptive decision-making through environment interaction, without relying on global network information.

\subsection{Symbiotic Communication for LAE}
\textcolor{black}{Symbiotic communication (SC) enables reciprocal collaboration among heterogeneous systems through service and resource exchange \cite{liang2022symbitoic}, where each system provides services or resources to others while receiving what it needs in return.} In LAE, satellite and terrestrial networks exhibit strong complementarity: satellites offer wide-area coverage for remote areas, while terrestrial networks provide low-latency and high-throughput links in urban regions. 
\textcolor{black}{Future research should investigate integrating satellite and terrestrial networks from an SC perspective to effectively leverage their complementary strengths and satisfy the diverse demands of LAVs.}

\textcolor{black}{\subsection{Resource Allocation in Satellite-Assisted LAE}
Efficient resource allocation is essential to enhance the performance of satellite-assisted LAE networks, where spectrum, beam directions, and satellite selection should be optimized under onboard resource constraints.
In this work, we adopt simple resource allocation schemes in a basic scenario.
Future works should address more complex scenarios involving diverse LAV demands, imperfect CSI, and Doppler effects. This calls for multi-objective optimization frameworks, robust optimization methods, and orthogonal time frequency space (OTFS) based resource allocation to enhance system performance}.



\section{Conclusion}\label{sec:conclu}
In this paper, we have investigated satellite-assisted LAE networks, where satellites with different orbital conditions and functions cooperate to serve LAVs in LAE.
First, we have presented an overview of LAE and satellites, including architecture, attributes, related standards, and existing systems.
Then, we have explored satellite assistance in LAE across three domains: communication, control, and computation.
Subsequently, we have proposed a framework for satellite-assisted LAE that leverages the distributed MIMO technique to form distributed LAV MIMO and distributed satellite MIMO.
The distributed MIMO introduces high channel diversity, enabling LAVs to reduce transmission power and satellites to extend service duration.
Moreover, we have developed a two-timescale optimization scheme to control beam directions and digital signal processing, reducing control signaling overhead. Our case study has confirmed that the proposed framework efficiently increases the energy efficiency of LAVs, reduces handovers, and balances performance and control overhead.


\bibliographystyle{IEEEtran}
\bibliography{IEEEabrv, sagin_mag}

\begin{thebibliography}{10}
\providecommand{\url}[1]{#1}
\csname url@samestyle\endcsname
\providecommand{\newblock}{\relax}
\providecommand{\bibinfo}[2]{#2}
\providecommand{\BIBentrySTDinterwordspacing}{\spaceskip=0pt\relax}
\providecommand{\BIBentryALTinterwordstretchfactor}{4}
\providecommand{\BIBentryALTinterwordspacing}{\spaceskip=\fontdimen2\font plus
\BIBentryALTinterwordstretchfactor\fontdimen3\font minus
  \fontdimen4\font\relax}
\providecommand{\BIBforeignlanguage}[2]{{%
\expandafter\ifx\csname l@#1\endcsname\relax
\typeout{** WARNING: IEEEtran.bst: No hyphenation pattern has been}%
\typeout{** loaded for the language `#1'. Using the pattern for}%
\typeout{** the default language instead.}%
\else
\language=\csname l@#1\endcsname
\fi
#2}}
\providecommand{\BIBdecl}{\relax}
\BIBdecl

\bibitem{jiang20236g}
Y.~Jiang, X.~Li, G.~Zhu, H.~Li, J.~Deng, K.~Han, C.~Shen, Q.~Shi, and R.~Zhang,
  ``{6G} non-terrestrial networks enabled low-altitude economy: Opportunities
  and challenges,'' \emph{arXiv preprint arXiv:2311.09047}, 2023.

\bibitem{baltaci2021survey}
A.~Baltaci, E.~Dinc, M.~Ozger, A.~Alabbasi, C.~Cavdar, and D.~Schupke, ``A
  survey of wireless networks for future aerial communications ({FACOM}),''
  \emph{{IEEE} Commun. Surveys Tuts.}, vol.~23, no.~4, pp. 2833--2884, Aug.
  2021.

\bibitem{zeng2019accessing}
Y.~Zeng, Q.~Wu, and R.~Zhang, ``Accessing from the sky: A tutorial on {UAV}
  communications for {5G} and beyond,'' \emph{Proc. {IEEE}}, vol. 107, no.~12,
  pp. 2327--2375, Dec. 2019.

\bibitem{zhu2018cooperative}
X.~Zhu, C.~Jiang, L.~Yin, L.~Kuang, N.~Ge, and J.~Lu, ``Cooperative multigroup
  multicast transmission in integrated terrestrial-satellite networks,''
  \emph{{IEEE} J. Sel. Areas Commun.}, vol.~36, no.~5, pp. 981--992, May 2018.

\bibitem{he2024toward}
S.~He, J.~Ge, Y.-C. Liang, and D.~Niyato, ``Toward symbiotic {STIN} through
  inter-operator resource and service sharing: Joint orchestration of user
  association and radio resources,'' \emph{{IEEE} J. Sel. Areas Commun.},
  vol.~42, no.~12, pp. 3674 -- 3689, Sep. 2024.

\bibitem{hu2020joint}
Y.~Hu, M.~Chen, and W.~Saad, ``Joint access and backhaul resource management in
  satellite-drone networks: A competitive market approach,'' \emph{{IEEE}
  Trans. Wireless Commun.}, vol.~19, no.~6, pp. 3908--3923, Jun. 2020.

\bibitem{wang2025toward}
Y.~Wang, G.~Sun, Z.~Sun, J.~Wang, J.~Li, C.~Zhao, J.~Wu, S.~Liang, M.~Yin,
  P.~Wang \emph{et~al.}, ``Toward realization of low-altitude economy networks:
  Core architecture, integrated technologies, and future directions,''
  \emph{arXiv preprint arXiv:2504.21583}, 2025.

\bibitem{tang2022gnss}
C.~Tang, Y.~Wang, L.~Zhang, and Y.~Zhang, ``{GNSS}/inertial navigation/wireless
  station fusion {UAV} {3-D} positioning algorithm with urban canyon
  environment,'' \emph{{IEEE} Sensors J.}, vol.~22, no.~19, pp.
  18\,771--18\,779, Aug. 2022.

\bibitem{leyva2023satellite}
I.~Leyva-Mayorga, M.~Martinez-Gost, M.~Moretti, A.~P{\'e}rez-Neira, M.~{\'A}.
  V{\'a}zquez, P.~Popovski, and B.~Soret, ``Satellite edge computing for
  real-time and very-high resolution earth observation,'' \emph{{IEEE} Trans.
  Commun.}, vol.~71, no.~10, pp. 6180--6194, Oct. 2023.

\bibitem{han2023two}
D.~Han, Q.~Ye, H.~Peng, W.~Wu, H.~Wu, W.~Liao, and X.~Shen, ``Two-timescale
  learning-based task offloading for remote {IoT} in integrated
  satellite-terrestrial networks,'' \emph{{IEEE Internet Things J.}}, vol.~10,
  no.~12, pp. 10\,131--10\,145, Jan. 2023.

\bibitem{fang2022olive}
Q.~Fang, Z.~Zhai, S.~Yu, Q.~Wu, X.~Gong, and X.~Chen, ``Olive branch learning:
  A topology-aware federated learning framework for space-air-ground integrated
  network,'' \emph{{IEEE} Trans. Wireless Commun.}, vol.~22, no.~7, pp.
  4534--4551, Dec. 2022.

\bibitem{heo2023mimo}
J.~Heo, S.~Sung, H.~Lee, I.~Hwang, and D.~Hong, ``{MIMO} satellite
  communication systems: A survey from the {PHY} layer perspective,''
  \emph{{IEEE} Commun. Surveys Tuts.}, vol.~25, no.~3, pp. 1543--1570, Jul.
  2023.

\bibitem{yuan2024cache}
S.~Yuan, Y.~Sun, and M.~Peng, ``Cache-aware cooperative multicast beamforming
  in dynamic satellite-terrestrial networks,'' \emph{{IEEE} Trans. Veh.
  Technol.}, pp. 1433--1445, Oct. 2024.

\bibitem{omid2023space}
Y.~Omid, Z.~M. Bakhsh, F.~Kayhan, Y.~Ma, and R.~Tafazolli, ``Space {MIMO}:
  Direct unmodified handheld to multi-satellite communication,'' in \emph{Proc.
  {IEEE} Global Commun. Conf. ({GLOBECOM})}, Kuala Lumpur, Malaysia, 2023, pp.
  1447--1452.

\bibitem{liang2022symbitoic}
Y.-C. Liang, R.~Long, Q.~Zhang, and D.~Niyato, ``Symbiotic communications:
  Where {Marconi} meets {Darwin},'' \emph{{IEEE} Wireless Commun. Mag.},
  vol.~29, no.~1, pp. 144--150, Feb. 2022.

\end{thebibliography}

\end{document}